\documentclass[letterpaper,aps,prb,citeautoscript,superscriptaddress,twocolumn]{revtex4-2} 
\usepackage[english]{babel}
\usepackage[utf8]{inputenc}

\usepackage{amsmath}
\usepackage{mathtools}
\usepackage{physics}
\usepackage{xcolor}
\usepackage{graphicx}
\usepackage[left=14mm,right=14mm,top=18mm,bottom=18mm,columnsep=15pt]{geometry} 
\usepackage{adjustbox}
\usepackage{placeins}
\usepackage[T1]{fontenc}
\usepackage{lipsum}
\usepackage{csquotes}

\usepackage{MnSymbol}
\usepackage{mathbbol}

\usepackage{graphicx}

\usepackage{bm}
\usepackage[
	hyperindex, 
	breaklinks, 
	colorlinks=true,
	hypertexnames=false]{hyperref}
\hypersetup{
    colorlinks   = true, 
    urlcolor     = blue, 
    linkcolor    = blue, 
    citecolor   = blue 
    }

\begin{document}
\title{Significant reduction in semiconductor interface resistance via interfacial atomic mixing}

\author{Qichen Song}
    \affiliation{Department of Mechanical Engineering, Massachusetts Institute of Technology, Cambridge, Massachusetts 02139} 
\author{Jiawei Zhou}
    \affiliation{Department of Mechanical Engineering, Massachusetts Institute of Technology, Cambridge, Massachusetts 02139} 
\author{Gang Chen}
    \email{gchen2@mit.edu}  
    \affiliation{Department of Mechanical Engineering, Massachusetts Institute of Technology, Cambridge, Massachusetts 02139}


\begin{abstract}
    The contact resistance between two dissimilar semiconductors is determined by the carrier transmission through their interface. Despite the ubiquitous presence of interfaces, quantitative simulation of charge transport across such interfaces is difficult, limiting the understanding of interfacial charge transport. This work employs Green's functions to study the charge transport across representative Si/Ge interfaces. For perfect interfaces, it is found that the transmittance is small and the contact resistance is high, not only because the mismatch of carrier pockets makes it hard to meet the momentum conservation requirement, but also because of the incompatible symmetries of the Bloch wave functions of the two sides. In contrast, atomic mixing at the interface increases the carrier transmittance as the interface roughness opens many nonspecular transmission channels, which greatly reduces the contact resistance compared with the perfect interface. Specifically, we show that disordered interfaces with certain symmetries create more nonspecular transmission. The insights from our study will benefit the future design of high-performance heterostructures with low contact resistance.

\end{abstract}


\maketitle

\section{Introduction} \label{sec:1}

The importance of interfaces in advanced semiconductor devices has been clearly pointed out
by Herbert Kroemer with his famous statement, ``the interface is the device''~\cite{RevModPhys.73.783}.
Semiconductor heterostructures play essential roles in vertical-cavity surface-emitting lasers~\cite{soda1979gainasp,tai1990drastic,peters1993band}, heterostructure bipolar transistors~\cite{kroemer1982heterostructure,lundstrom1983numerical,jain1991structure}, quantum cascade lasers~\cite{page2001300}, quantum well infrared photodetectors~\cite{levine1990high}, thermionic microcoolers~\cite{chen2000energy,zeng1999sige}, spin qubit  devices~\cite{scappucci2020germanium,PhysRevB.103.125201}, thermoelectric power generators~\cite{Hinsche_2012,doi:10.1063/1.125040,koga2000experimental,PhysRevLett.92.106103}, \textit{etc}. 
However, the interfaces in heterostructures strongly scatter electrons and 
cause the contact resistance~\cite{landauer1957spatial,PhysRevB.49.17177,chen2003diffusion,schroeder1994modelling}. The interface scattering probabilities are not only determined by the intrinsic properties of bulk materials, but by the non-intrinsic properties such as the interface structures. Specifically, the interface roughness due to atomic mixing~\cite{takeuchi2002observation}, as a common type of interface disorder, alters the contact resistance.
In order to design proper interface structures that minimize interfacial resistance, it is crucial to understand how the atomic mixing affects electron scattering at interfaces. 

The nonequilibrium Green's function (NEGF) is often used to describe the structure-dependent charge transport~\cite{datta1997electronic}. 
Many works using NEGF combined with Landauer formula for conductance are conducted to study the transport across molecular junctions~\cite{PhysRevB.69.035108,PAPIOR20178,PhysRevB.76.115117,PhysRevB.63.245407}, nanotransistors~\cite{datta2000nanoscale,rahman2003theory}, grain boundaries in two-dimensional materials~\cite{yazyev2010electronic}, metal-semiconductor interfaces~\cite{PhysRevB.93.155302,PhysRevApplied.10.024016}, metal-metal interfaces in magnetic multilayers ~\cite{PhysRevLett.69.1676,PhysRevB.42.8110,PhysRevB.51.283,PhysRevB.55.960} and semiconductor interfaces~\cite{PhysRevApplied.16.054028,PhysRevApplied.14.024037}.
In particular, Bellotti \textit{et al.} investigated the carrier transport through semiconductor interfaces in the presence of positional and compositional disorders using NEGF and found that the disorder significantly impedes the coherent propagation of carriers through multiple interfaces~\cite{PhysRevApplied.16.054028}. Tibaldi \textit{et al.} performed a large-scale NEGF calculation of the carrier transport in a realistic tunnel junctions for vertical-cavity surface-emitting lasers and achieved a good agreement with experimental $I$-$V$ curve. 
However, the interface roughness in the transverse directions is neglected in these works, as the computational cost of NEGF increases dramatically with the cross-section areas of the interface.
Aside from NEGF calculation, 
Daryoosh \textit{et al.} used a simple effective mass model to study the carrier transport through barriers in metal-based superlattices and found that the nonspecular (diffuse) scattering can dramatically increase the thermoelectric figure of merit $zT$~\cite{PhysRevLett.92.106103}.
Los studied how the transmission probability varies with the average fluctuations of potential energies due to interface disorders under the effective mass approximation~\cite{PhysRevB.72.115441}. However, the effective mass approximation adopted in these works can poorly describe practical semiconductors with band pockets not at zone center. 
Due to the multi-valley nature~\cite{doi:10.1063/1.125040} of the band structures of semiconductors, new physics shall emerge for electron interfacial transport.

\begin{figure*}[t]
    \includegraphics[width=\textwidth]{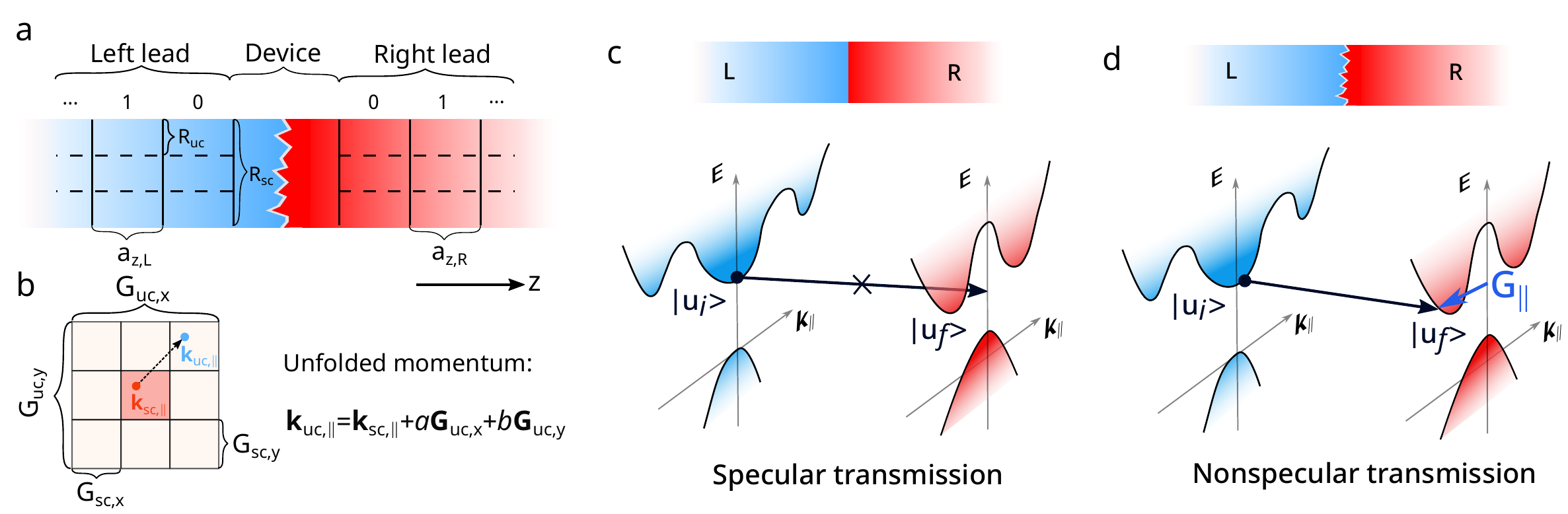}
    \caption{(a) The partitioning for Green's function calculation with repeated unit cell in the lead region being numbered with 0, 1 .... (b) The transverse momentum conservation law and how the transverse momenta in the unit-cell representation and the supercell representation are related. (c) Specular transmission process through a perfect interface where the transverse momentum is conserved, $\mathbf{k}_{i,\parallel} = \mathbf{k}_{f,\parallel}$. (d)  Nonspecular transmission processes through a disordered interface, where $\mathbf{k}_{i,\parallel} = \mathbf{k}_{f,\parallel} +m\mathbf{G}_x+n\mathbf{G}_y$, allows the scattering between valleys with different $\mathbf{k}_\parallel$.}
    \label{gf}
\end{figure*}
 
In this work, we apply the mode-resolved Green's function formalism with tight-binding Hamiltonian to study charge transport across perfect and rough interfaces due to atomic mixing.
In particular, we take advantage of the transverse translational symmetry to reduce the computational cost of surface Green's function. The tight-binding Hamiltonian makes sure the multiple carrier pockets in the Brillouin zone are properly described.
We vary the degree of disorders in transverse directions and perpendicular direction and study the specular and nonspecular interface scattering processes with mode resolution. Moreover, we unveil the roles of disorders and  symmetries in assisting nonspecular transmission. We show that over one order of magnitude of reduction of the specific contact resistance can be achieved by the interfacial atomic mixing.  

\section{Methodology}

In the Green's function calculation, we first divide the system into three regions, two semi-infinite lead regions and a device region, as depicted in Fig.~\ref{gf} (a). The repeated cells along $z$ direction in the lead region are indexed by 0, 1, ... and the period length is $a_{z,\alpha}$ with $\alpha = L,R$. The whole supercell structure is periodic along directions parallel to the interface. 
Inside the lead region, there are $N_{\mathrm{uc},x}\times N_{\mathrm{uc},y}$ identical unit cells along the transverse directions. The transverse lattice vector for the supercell is $\mathbf{R}_{\mathrm{sc},\beta} = N_{\mathrm{uc},\beta} \mathbf{R}_{\mathrm{uc},\beta}$ with $\beta=x,y$, where $\mathbf{R}_{\mathrm{uc},\beta}$ is the transverse lattice vector for the unit cell.
As a result, the transverse momentum in the supercell representation can be uniquely unfolded in a momentum defined in the unit-cell representation, as  elucidated in Fig.~\ref{gf} (b). The unfolded momentum can be expressed by,
 \begin{equation}
     \mathbf{k}_{\mathrm{uc},\parallel} = \mathbf{k}_{\mathrm{sc},\parallel} + a\mathbf{G}_{\mathrm{sc},x}+b\mathbf{G}_{\mathrm{sc},y}
 \end{equation}
where $\mathbf{G}_{\mathrm{sc},x}$ and $\mathbf{G}_{\mathrm{sc},y}$ are the transverse reciprocal lattice vectorsa, and $a$ and $b$ are integers to be determined. Finding the correct pair of $a$ and $b$ is known as an unfolding problem and we use the unfolding scheme by Popescu and Zunger~\cite{PhysRevLett.104.236403} to resolve the correct $\mathbf{k}_{\mathrm{uc},\parallel}$. 



We consider the elastic interface scattering limit, where the energy $E$ of the incident electron is conserved. In addition,
the in-plane translational symmetry of the supercell dictates that the transverse momentum $\mathbf{k}_{\mathrm{sc},\parallel}$ must be conserved during an interface scattering event. When the device region contains a perfect interface with the same in-plane periodicity as the lead region, the transverse momentum $\mathbf{k}_{\mathrm{uc},\parallel}$ is conserved. However, when the device region consists of a rough interface, $\mathbf{k}_{\mathrm{uc},\parallel}$ is not always conserved. This is because the interface roughness breaks the internal transverse translational symmetry within the supercell and $\mathbf{k}_{\mathrm{sc},\parallel}$ can be unfolded into different $\mathbf{k}_{\mathrm{uc},\parallel}$ for the incident state and the transmitted state. As illustrated in Fig.~\ref{gf} (c), for a perfect interface, $\mathbf{k}_{\mathrm{uc},\parallel}$ is conserved, and we denote this type of scattering process the specular transmission. For a rough interface shown in Fig.~\ref{gf} (d),  $\mathbf{k}_{\mathrm{uc},\parallel}$ can be either conserved or nonconserved. Particularly, we denote the scattering process with nonconserved $\mathbf{k}_{\mathrm{uc},\parallel}$ the nonspecular transmission.

We define the transmission probability matrix from the left side $T_{ji}(E,\mathbf{k}_{\mathrm{sc},\parallel})$ as the ratio between the normal current of the transmitted state $j$ to the incident state $i$. 
 Formally, we can express the specular and nonspecular transmission probability matrix with,
\begin{equation}
\begin{cases}
T_{\mathrm{s},ji}(E,\mathbf{k}_{\mathrm{sc},\parallel})=T_{ji}(E,\mathbf{k}_{\mathrm{sc},\parallel}), \;\text{when}\; \mathbf{k}_{\mathrm{uc},\parallel,j}=\mathbf{k}_{\mathrm{uc},\parallel,i}\\
 T_{\mathrm{ns},ji}(E,\mathbf{k}_{\mathrm{sc},\parallel})=T_{ji}(E,\mathbf{k}_{\mathrm{sc},\parallel}), \;\text{when}\;  \mathbf{k}_{\mathrm{uc},\parallel,j}\neq\mathbf{k}_{\mathrm{uc},\parallel,i}
\end{cases}
\end{equation}
The elements of the transmission probability matrix from the left side is given by,
\begin{equation}
     T_{ji}(E,\mathbf{k}_{\mathrm{sc},\parallel}) = |t_{RL,ji}(E,\mathbf{k}_{\mathrm{sc},\parallel})|^2
\end{equation}
where the transmission matrix $t_{RL,ji}(E,\mathbf{k}_{\mathrm{sc},\parallel})$ is related to the Green's function via the following relation~\cite{PhysRevB.72.035450}:
\begin{equation}
     t_{RL}(E,\mathbf{k}_{\mathrm{sc},\parallel}) = i\sqrt{V^r_R}[U^r_R]^{-1}G_{N+1,0}[U^{a\dagger}_L]^{-1}\sqrt{V^a_L}
     \label{trl}
\end{equation}
The formal definitions and detailed calculations of the velocity matrices $V_{R/L}^{r/a}$, eigenvector matrices  $U_{R/L}^{r/a}$,
 and Green's function $G_{N+1,0}$ can be found in Appendix~\ref{mgf1}. Note that the calculation of the velocity matrices and eigenvector matrices require the surface Green's function $g^{a/r}_{L/R}(E,\mathbf{k}_{\mathrm{sc},\parallel})$. We apply the Fourier transform to the Hamiltonian to obtain the block-diagonal surface Green's function. Then, we apply the inverse Fourier transform to obtain the surface Green's function. These procedures allow us to invert small matrix multiple times rather than directly inverting the large matrix, which greatly boosts the computational efficiency. The detailed implementation can be found in our prior work on studying diffuse phonon scattering by rough interfaces~\cite{PhysRevB.104.085310}.

\begin{figure*}[t]
\centering
    \includegraphics[width=0.95\textwidth]{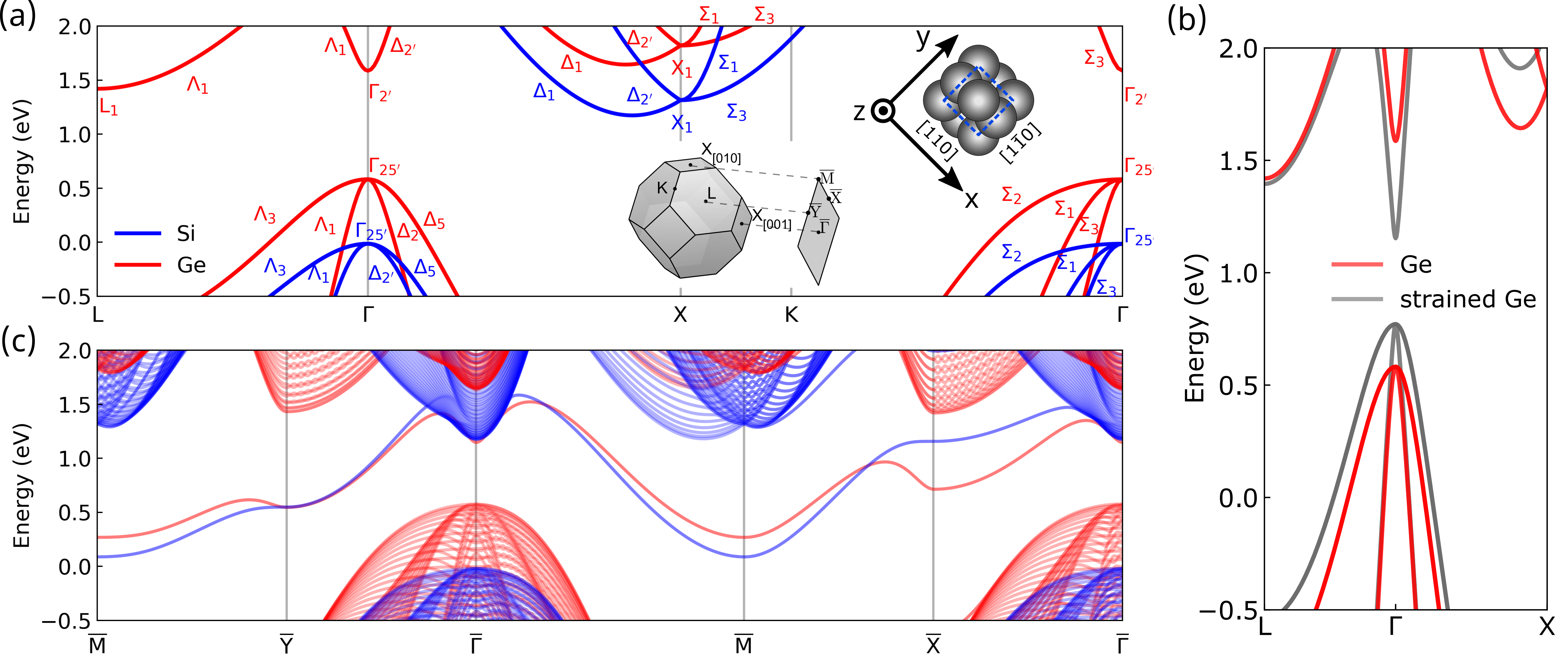}
    \caption{(a) The band structure for bulk Si, Ge along high-symmetry lines in the first Brillouin zone. Left inset: the three-dimensional first Brillouin zone of Si (Ge) and its projection on the (001) plane. Right inset: the atomic structure for Si (Ge) unit cell along [001] direction.  (b) The band structures for Ge and tensile-strained Ge (2 \%). (c) The band structures for Si and Ge slabs along high-symmetry lines in the surface Brillouin zone. The slab contains 108 atom layers (27$a$ in thickness with $a$ the lattice constant). }
    \label{bulk}
\end{figure*}


\section{Band structures}\label{struc}

We study the [001] Si/Ge interface as it is a classical semiconductor interface used in a wide range of applications such as quantum information storage~\cite{PhysRevB.103.125201}, strained field-effect transistors~\cite{xia2007si,hashemi2011high} and thermoelectrics~\cite{yang2002measurements,doi:10.1063/1.125040}. 
To start with, we examine the bands structures for Si and Ge individually.
We use $sp^3d^5s^*$ Slater-Koster tight-binding model~\cite{PhysRevB.57.6493,PhysRev.94.1498} to construct the Hamiltonian, with hopping integral parametrizations from Niquet \textit{et al}~\cite{PhysRevB.79.245201}. More specifically, the hopping integral varies with the bond length according to a power law,
\begin{equation}
V_{\alpha\beta\gamma}(L)=V_{0,\alpha\beta\gamma}\left(\frac{L}{L_0}\right)^{\chi_{\alpha\beta\gamma}}
\label{vhop}
\end{equation}
where $\alpha$ and $\beta$ refer to the orbital types, $\gamma$ is the type of bond, $L$ is the bond length, $L_0$ is the unstrained bond length, $V_{0,\alpha\beta\gamma}$ is the hopping integral for unstrained bond, and $\chi_{\alpha\beta\gamma}$ is the power-law exponent. The band structures using this set of hopping integrals have shown an excellent agreement with \textit{GW} calculations with various strain ratios~\cite{PhysRevB.79.245201}. 
The bulk Si and Ge have mismatched lattice constants with $a_{\mathrm{Si}} = 5.431$ \AA $\,$ and $a_{\mathrm{Ge}} = 5.658$ \AA. Correspondingly, the unstrained bond lengths for Si and Ge are 2.352 and 2.450 \AA $\,$, respectively. For simplicity, we study the lattice-matched interface and we assume the Si-Si and Ge-Ge bond lengths are the same, $L=$  2.398 \AA$\,$, which is relaxed Si-Ge bond length found by Niquet \textit{et al}~\cite{PhysRevB.79.245201}. Furthermore, we rescale the $L_0$ for Si-Si bond and Ge-Ge bond to be  $L_0=$  2.398 \AA$\,$ to ensure that the Si's and Ge's band structures are the same with their unstrained bulk band structures~\cite{bond,offset}. In our calculation, the spin-orbital coupling is not included.




First, we compare the band structures $E_n(\mathbf{k})$ of bulk Si and Ge along high-symmetry paths and examine the distributions of electron/hole pockets in the first Brillouin zone, as depicted in Fig.~\ref{bulk} (a), which clearly shows that the conduction band pockets for Si and Ge are distributed very differently, whereas their valence band pockets are quite similar.  In particular, the highest valence bands for Si and Ge are both at $\Gamma$ point. In contrast, the conduction band edge for Si is close to the X point along the $\Gamma$X path (in the following, we denote this point $\Delta$), while the conduction band edge for Ge is at the L point. In addition, there are six pockets for Si's lowest conduction band at the $\Delta$ point, while there are four pockets (or eight half-pockets) for Ge at the L point.
The second-lowest conduction band for Si is at the X point. The second- and third-lowest conduction bands for Ge are at the $\Gamma$ point and $\Delta$ point, respectively.

Next, we look into the symmetry properties of the Bloch wave functions in order to develop an understanding of how symmetry affects the transmission. In the bra-ket notation, the transmission matrix element is directly proportional to the Green's function matrix element, 
\begin{equation}
    t_{ji}\propto\bra{u_j}\hat{G}\ket{u_i}
\end{equation}
where $\hat{G}=\left(EI-\hat{H}\right)^{-1}$ is the Green's function operator and $\hat{H}$ is the Hamiltonian operator~\cite{PhysRevB.23.6851,economou2006green}. It is easy to show that $\hat{G}$ inherits all symmetries of $\hat{H}$~\cite{PhysRevB.83.085426}. For $\ket{u_j}$ and $\ket{u_i}$ with certain type of symmetries, the transmission matrix element $t_{ji}$ is guaranteed to vanish according to group theory~\cite{dresselhaus2007group}. Thus, it is essential to identify the symmetries of Bloch wave functions of the two sides.

To describe the symmetry properties of Bloch wave functions in Si and Ge,
the Bouckaert-Smoluchowski-Wigner (BSW)~\cite{PhysRev.50.58} notation is
adopted in Fig.~\ref{bulk} (a), which marks the irreducible representations for the Bloch wave function. The different irreducible representations of the same group (labeled by the same Greek letter with different subscripts) are orthogonal to each other. The character tables for different groups can be found in group theory textbooks~\cite{dresselhaus2007group} and online databases~\cite{aroyo2006bilbao2}. They describe how the Bloch wave function transforms under different symmetry operations. For instance, the states of the lowest conduction band of Si at the $\Delta$ point transform as $\Delta_1$  representation under the symmetry operations of the $C_{4v}$ group. On the other hand, the states of the second-lowest conduction band in Ge at the $\Delta$ point transform as $\Delta_{2^\prime}$ representation. 
Without the loss of generality, we consider the $\Delta$ points along the \textit{z} axis [(0,0,1) axis]. In this case, one of the $C_{4v}$ group elements is the symmetry operation $\hat{S} = \{C_4|\bm{\tau}_d\}$ with $\bm{\tau}_d  = \frac{1}{4}(a,a,a)$, which first rotates the Bloch wave function by 90$^{\circ}$ with respect to the \textit{z} axis and then applies the translation operator by $\bm{\tau}_d$. When applying $\hat{S}$ to a state $\ket{u_{i}}$ of $\Delta_1$ symmetry, we have $\hat{S} \ket{u_{i}} = 1\cdot e^{ik_za/4}\ket{u_{i}}$, where $k_z$ is the wave vector's \textit{z} component. The phase factor $e^{ik_za/4}$ appears because the space group of Si (Ge) structure is nonsymmorphic. In comparison, when applying the same operator $\hat{S}$ to a state $\ket{u_{i}}$ of $\Delta_{2^\prime}$ symmetry, we have $\hat{S} \ket{u_{i}} = -1\cdot e^{ik_za/4}\ket{u_{i}}$. Intuitively, one can regard $\Delta_1$ as ``even'' and $\Delta_{2^\prime}$ as ``odd'' in a more generalized way. If the incident and transmitted states are not compatible, they will never interact. Hence, knowing the symmetry properties of wave functions (\textit{i.e.}, their irreducible representations) will be useful in the later analysis of the transmission probabilities.

\begin{figure*}[t]
    \includegraphics[width=0.92\textwidth]{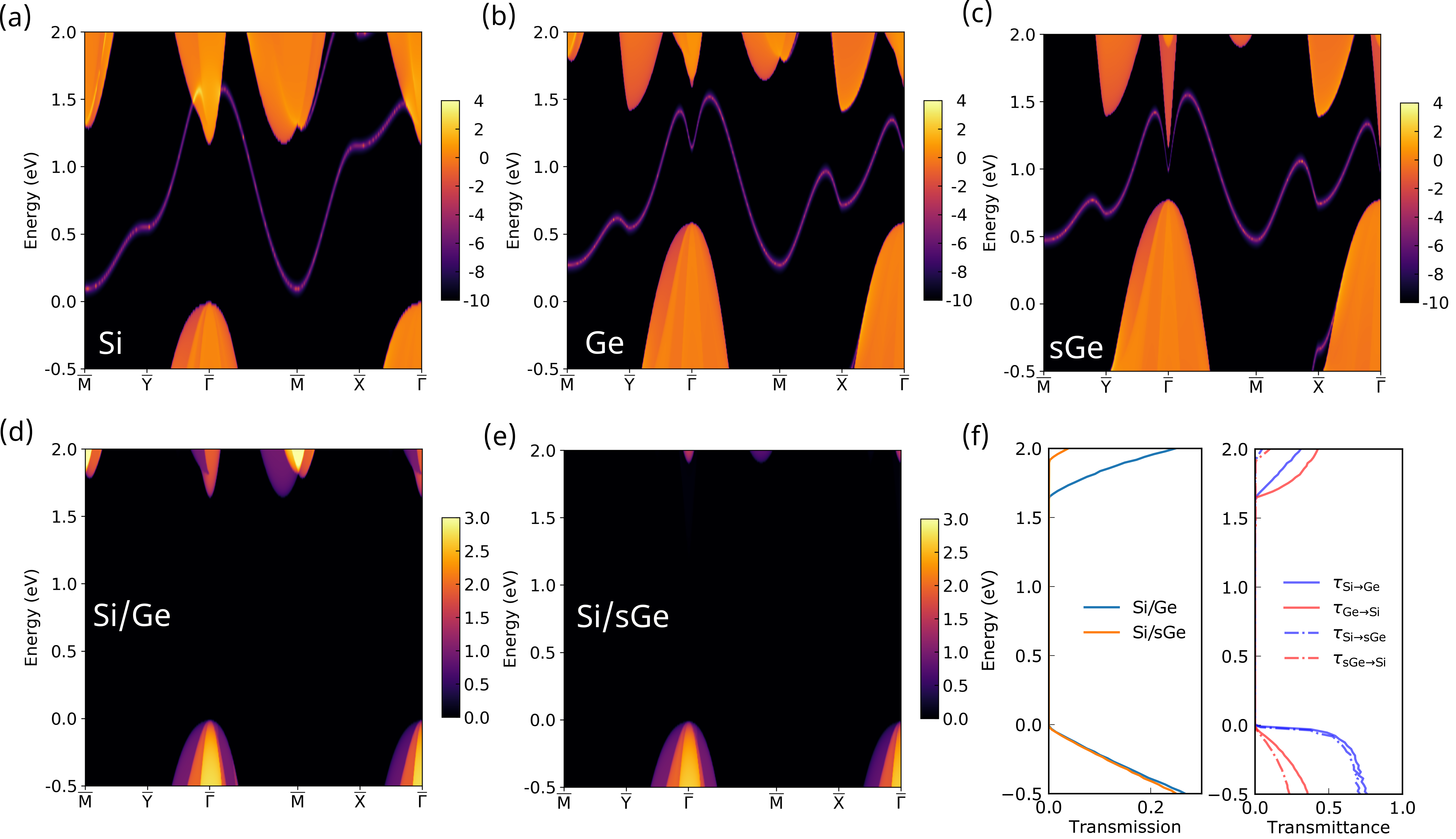}
    \caption{The surface density of states $\mathrm{SDOS}(E,\mathbf{k}_{\mathrm{uc},\parallel})$ for (a) Si, (b) Ge, and (c) sGe. The logarithm of the surface density of state is indicated by colors and the spin degree of freedom 2 is not multiplied. The transmission spectra $T(E,\mathbf{k}_{\mathrm{uc},\parallel})$ along high-symmetry lines in the surface Brillouin zone through (d) a perfect Si/Ge [001] interface and (e) a perfect Si/sGe [001] interface. The color indicates the value of $T(E,\mathbf{k}_{\mathrm{uc},\parallel})$. (f) Left panel: the transmission function $\Theta(E)=\frac{1}{N_{\mathbf{k}_{\mathrm{uc},\parallel}}}\sum_{\mathbf{k}_{\mathrm{uc},\parallel}}T(E,\mathbf{k}_{\mathrm{uc},\parallel})$ for a perfect Si/Ge  interface and a perfect Si/sGe interface ($N_{\mathbf{k}_{\mathrm{uc},\parallel}}=40\times40$ is used).  Right panel: the transmittance $\tau_\alpha(E)=\frac{\Theta(E)}{\Theta_{\mathrm{bulk},\alpha}(E)}$  from Si and Ge side, where $\Theta_{\mathrm{bulk},\alpha}(E)$ is the transmission function for bulk $\alpha$ material.}
    \label{sdos}
\end{figure*}

Moreover, the strain effect can change the relative positions for different valleys in the reciprocal space. We find that the strain generally has a smaller impact on Si compared with Ge, thus, we only consider the case of applying strain to Ge. Since we have already assumed the Si and Ge have the same bond lengths $L=$ 2.398 \AA, we change the equilibrium $L_0$ of Ge from 2.398 \AA$\,$ to  2.343 \AA$\,$ while keeping $L$ unchanged. The corresponding hopping integrals $V_{\alpha\beta\gamma}$ defined in Eq.~\ref{vhop} are altered and the Ge band structure is accordingly changed. Equivalently, we have applied a tensile strain of 2 \% to Ge. In Fig.~\ref{bulk} (b), we compare the band structures of strained Ge (sGe for short) with the relaxed Ge. 
The elongated bond pushes the second-lowest conduction band downwards and makes it the lowest conduction band. It also shifts the third-lowest band further upwards. Meanwhile, the strained Ge-Ge bond also makes the valence band upwards, thus causing a smaller band gap. The features of the sGe band structure are consistent with other works~\cite{PhysRevB.103.125201,michel2010high}.

When forming an interface, the translational symmetry is broken along the direction normal to the interface, and the band structures are now projected to the two-dimensional surface Brillouin zone, as depicted in the left inset of Fig.~\ref{bulk} (c). We conduct a slab calculation to study the projected band structure. The slab is periodic along the \textit{x} and \textit{y} directions and finite in the \textit{z} direction. The unit cell for Si (Ge) slab along [001] direction contains four atoms, as shown in the right inset of 
Fig.~\ref{bulk} (a). Note that the atomic structure of Si (Ge) has mirror symmetries with respect to (110) and (1$\bar{1}$0) planes.  In Fig.~\ref{bulk} (c), we find that the highest valence bands of Si and Ge are both projected to the $\bar{\Gamma}$ point. Two out of the six lowest conduction band pockets of Si are projected to the $\bar{\Gamma}$ point, two pockets are projected to a point between the $\bar{\Gamma}$ and $\mathrm{\bar{X}}$ points, and the remaining two pockets are projected to a point between the $\bar{\Gamma}$ and $\mathrm{\bar{Y}}$ points. As for Ge, two of the four lowest conduction band pockets at L points are projected to the $\mathrm{\bar{X}}$ point, and the remaining two are projected to the $\mathrm{\bar{Y}}$ point. Since we use a slab to compute the projected band structure, we observe the surface states~\cite{PhysRevB.89.115318} for Si and Ge in the band gap. They each have two degenerate surface states within the \textit{x}-\textit{y} plane, one for the top surface, one for the bottom surface. However, in the direction normal to the interface (\textit{z} direction), these surface states are localized, thus do not contribute to the interfacial transport.

Lastly, we study the density of states for the projected band. The density of states for the projected band structures at the given energy $E$ and transverse momentum $\mathbf{k}_{\mathrm{uc},\parallel}$ is obtained by taking the imaginary part of retarded surface Green's function given by Eq.~\ref{surfaceG} for the lead,
\begin{equation}
\mathrm{SDOS}(E,\mathbf{k}_{\mathrm{uc},\parallel})=-\frac{1}{\pi}\mathrm{Im} g^{r}_\alpha(E,\mathbf{k}_{\mathrm{uc},\parallel})
\label{imsg}
\end{equation}
where $g^r_\alpha(E,\mathbf{k}_{\mathrm{uc},\parallel})$ is the retarded surface Green's function for $\alpha$ lead  with $\alpha=\mathrm{Si,Ge}$.
From the density of states shown in Fig.~\ref{sdos} (a), where we use color to indicate $\mathrm{ln}\left[\mathrm{SDOS}(E,\mathbf{k}_{\mathrm{uc},\parallel})\right]$, we identify the localized states in the band gap, the continuum spectrum of propagating conduction band electrons and the resonant states inside the continuum spectrum. 

\begin{figure*}[t]
    \includegraphics[width=\textwidth]{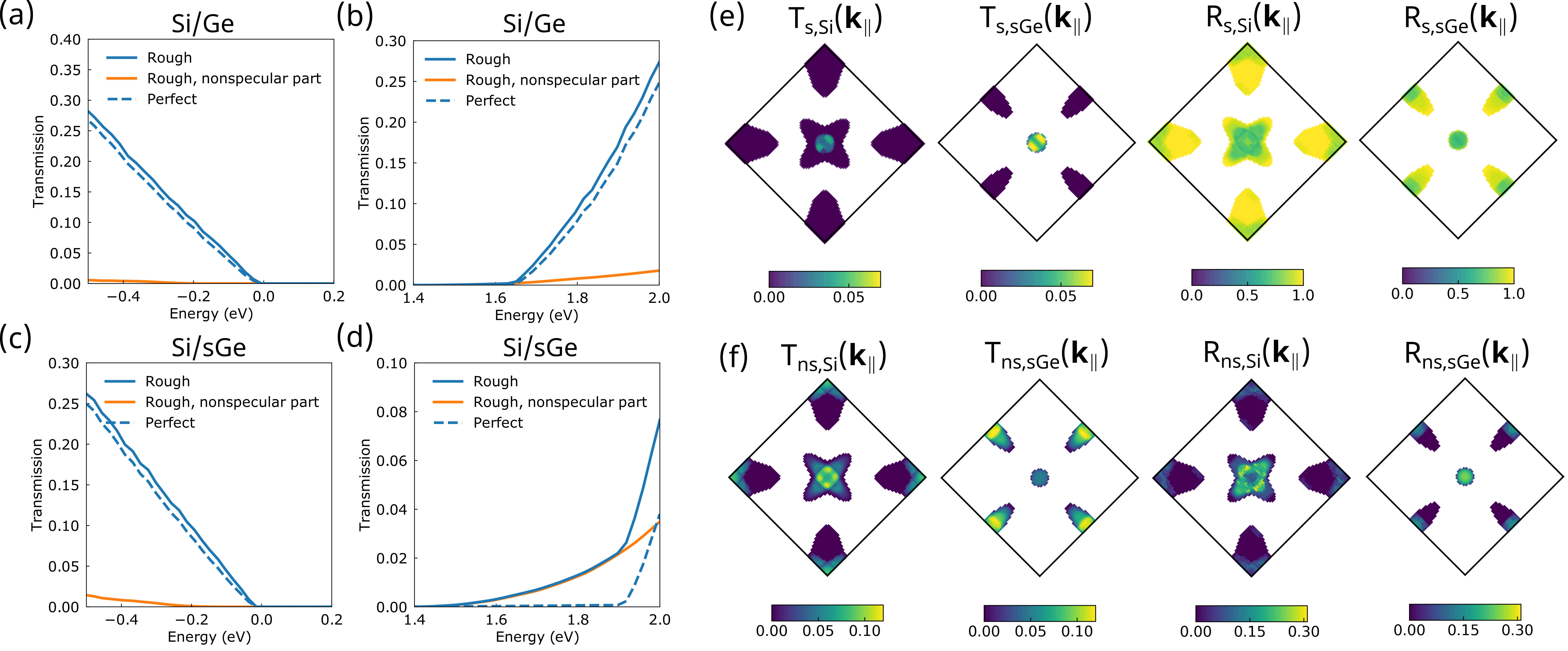}
    \caption{The ensemble-averaged total transmission function $\Theta(E)$ and nonspecular transmission function $\Theta_\mathrm{ns}(E)$ through rough Si/Ge interfaces for (a) electrons and (b) holes in solid lines compared with the transmission function for a perfect interface in dashed line. 21 configurations of $2\times2$, 8 ml disordered interfaces are used for ensemble average. (c), (d) The transmission function for Si/sGe interfaces. (e), (f) The ensemble-averaged mode-resolved specular and nonspecular scattering probabilities at $E$ = 1.77 eV for rough Si/sGe interfaces defined by Eq.~\ref{mode_total} as a function of in-plane momentum $\mathbf{k}_{\mathrm{uc},\parallel}$. The calculation uses a 40 $\times$ 40 $\mathbf{k}_{\mathrm{sc},\parallel}$-point mesh. Equivalently, it corresponds to an 80 $\times$ 80 $\mathbf{k}_{\mathrm{uc},\parallel}$ mesh.}
    \label{ns}
\end{figure*}

\section{Transmission through a perfect interface}\label{perfecti}
We first study the electron transmission through a perfect interface. There are several relevant physical quantities, and we want to clarify their definitions here to avoid confusion. 
$t_{RL,ji}(E,\mathbf{k}_{\mathrm{sc},\parallel})$ is the transmission matrix, which is used to compute the transmission probability matrix. $T_{ji}(E,\mathbf{k}_{\mathrm{sc},\parallel})$ is the transmission probability matrix, which described mode-to-mode transition probability, normalized by normal incident current. $T_i(E,\mathbf{k}_{\mathrm{sc},\parallel})$ refers the transmission probability (transmittance) across the interface for an incident electron $i$. The transmission spectrum $T(E,\mathbf{k}_{\mathrm{sc},\parallel})$ is the number of transmission channels including all subbands that have same $E$ and $\mathbf{k}_{\mathrm{sc},\parallel}$. The transmission function (we use transmission in short in figures) $\Theta(E)$ describes the total number of transmission channels at the given energy $E$ and is the sum of all transmission channels with different $\mathbf{k}_{\mathrm{sc},\parallel}$. Note that $t_{RL,ji}(E,\mathbf{k}_{\mathrm{sc},\parallel})$, $T_{ji}(E,\mathbf{k}_{\mathrm{sc},\parallel})$,  $T_i(E,\mathbf{k}_{\mathrm{sc},\parallel})$ depend on which side incident electron is from, whereas $T(E,\mathbf{k}_{\mathrm{sc},\parallel})$, $\Theta(E)$ are independent of the side of incidence.

The transmission probability can be computed by summing the transmission probability matrix over all possible final states. The specular part and nonspecular part of the transmission probability read as,
\begin{equation}
 \begin{cases}
 T_{\mathrm{s},i}(E,\mathbf{k}_{\mathrm{sc},\parallel})=\sum_j T_{\mathrm{s},ji}(E,\mathbf{k}_{\mathrm{sc},\parallel})\\
 T_{\mathrm{ns},i}(E,\mathbf{k}_{\mathrm{sc},\parallel})=\sum_j T_{\mathrm{ns},ji}(E,\mathbf{k}_{\mathrm{sc},\parallel})
 \end{cases}
 \label{mode_total}
\end{equation}
The transmission function is the measure of conductance channels and  can be expressed by,
$\Theta(E) = \Theta_\mathrm{s}(E)+\Theta_\mathrm{ns}(E)$, where
the specular and nonspecular transmission functions are defined by,
 \begin{equation}
     \Theta_\mathrm{s}(E) = \frac{1}{N_{\mathbf{k}_{\mathrm{sc},\parallel}}}\sum_{i,\mathbf{k}_{\mathrm{sc},\parallel}}T_{\mathrm{s},i}(E,\mathbf{k}_{\mathrm{sc},\parallel})
\end{equation}
\begin{equation}
     \Theta_\mathrm{ns}(E) = \frac{1}{N_{\mathbf{k}_{\mathrm{sc},\parallel}}}\sum_{i,\mathbf{k}_{\mathrm{sc},\parallel}}T_{\mathrm{ns},i}(E,\mathbf{k}_{\mathrm{sc},\parallel})
\end{equation}
For the case of perfect interface, all the transmission processes are specular, hence we have $T_{\mathrm{s},ji}(E,\mathbf{k}_{\mathrm{sc},\parallel})=T_{ji}(E,\mathbf{k}_{\mathrm{sc},\parallel})$. In addition, for the perfect interface, we only need to construct a unit cell as the supercell such that the in-plane momenta in the unit-cell representation and the supercell representation are the same, $\mathbf{k}_{\mathrm{uc},\parallel}=\mathbf{k}_{\mathrm{sc},\parallel}$.


The transmission spectrum is attained by $T(E,\mathbf{k}_{\mathrm{uc},\parallel})=\sum_i T_i(E,\mathbf{k}_{\mathrm{uc},\parallel})$, where we sum over all subbands with the same $E$ and $\mathbf{k}_{\mathrm{uc},\parallel}$. In Figs.~\ref{sdos} (d) and (e), we show the transmission spectra $T(E,\mathbf{k}_{\mathrm{uc},\parallel})$ through the Si/Ge and Si/sGe interfaces. Comparing with the surface density of states through examining the Figs.~\ref{sdos} (a)-(d), we see that the transmission is non-zero only when the surface density of states for Si and sGe overlap. This is due to the energy and momentum conservation requirement. 
For example, the Ge's lowest conduction band at $\mathrm{\bar{X}}$ and $\mathrm{\bar{Y}}$ does not have any corresponding states in Si, thus cannot contribute to transmission. Most of the overlapped states are the valleys at $\mathrm{\bar{\Gamma}}$ and along the $\mathrm{\bar{\Gamma}}\mathrm{\bar{M}}$ path, which corresponds to the lowest conduction band in Si, and second- and third-lowest conduction bands in Ge.

Because of the mismatch of conduction band valleys of Si and Ge, a large ``transport gap'' of 1.65 eV emerges at the $\mathrm{\bar{\Gamma}}$ point. For a Si/sGe interface, the transmission spectra for holes change slightly from a Si/Ge interface. The transport gap is 1.91 eV, which is even larger due to fewer energy and momentum matched conduction bands. From the energy-resolved transmission and transmittance in Fig.~\ref{sdos} (f), we also find that strain has much smaller impact on the hole transmission than the electron transmission. This is because the valence bands stay at the $\mathrm{\bar{\Gamma}}$ point even with strain, while the strain changes the position of conduction bands in reciprocal space more profoundly.

What is intriguing is that at the $\mathrm{\bar{\Gamma}}$ point, Si and sGe have overlapped conduction band pockets, yet the transmission $T(E,\mathbf{k}_{\mathrm{uc},\parallel})$ is still almost zero. This implies that there are other factors other than energy and momentum conservation which limits the transmission. We found out that the zero transmission originates from the different symmetries of the wave functions. In three-dimensional Brillouin zone, the lowest conduction band of sGe is at the $\Gamma$ point with $\Gamma_{2^\prime}$ symmetry. Under the symmetry operation $\hat{S} = \{C_4|\bm{\tau}_d\}$ mentioned above, it transforms as $\hat{S}\ket{u_{R,\Gamma}} = - \ket{u_{R,\Gamma}}$.  
In comparison, for the lowest conduction band of Si at the $\Delta$ point, it satisfies $\hat{S}\ket{u_{L,\Delta}} = e^{ik_{L,z}a/4}\ket{u_{L,\Delta}}$. The Hamiltonian for a perfect Si/sGe interface should always have ``even'' symmetry representation $\Delta_1$. Hence, it follows that $\hat{S}\hat{H}=e^{ik_{L,z}a/4}\hat{H}$ and $\hat{S}\hat{G}=e^{-ik_{L,z}a/4}\hat{G}$. As a result, the transmission matrix element should satisfy the condition,
$t_{ji} \propto \bra{u_{R,\Gamma}}\hat{G}\ket{u_{L,\Delta}} = \bra{\hat{S}u_{R,\Gamma}}\hat{S}\hat{G}\ket{\hat{S}u_{L,\Delta}} = -\bra{u_{R,\Gamma}}\hat{G}\ket{u_{L,\Delta}}$. Consequently, we obtain that $T_{ji} =|t_{ji}|^2 = 0$. 
Similarly, for the electrons at the $\Delta$ point with $\mathbf{k}=(0,0,k_{R,z})$, they have $\Delta_{2^\prime}$ symmetry and transform as $\hat{S}\ket{u_{R,\Delta}} = - e^{i k_{R,z} a/4}\ket{u_{R,\Delta}}$. Resultantly, we have $t_{ji} \propto \bra{u_{R,\Delta}}\hat{G}\ket{u_{L,\Delta}} = \bra{\hat{S}u_{R,\Delta}}\hat{S}\hat{G}\ket{\hat{S}u_{L,\Delta}} = -\bra{u_{R,\Delta}}\hat{G}\ket{u_{L,\Delta}}$ and,  correspondingly, $T_{ji} = 0$. In short, the transmission at $\bar{\Gamma}$ is exactly zero, dictated by symmetry.

\section{Transmission through rough interfaces}\label{roughi}

We add interface disorders in the form of atomic mixing. In particular, we randomly swap pairs of Si and Ge atoms that have the same distance to the interface. We use a larger supercell with in-plane periodicity to describe the rough interface. To mimic an actual rough Si/Ge interface observed in experiments~\cite{takeuchi2002observation}, we make sure that the further away from the interface, the fewer or equal number of atom pairs are swapped. 
In the following, we define two measures of the degree of interface disorders along the interface normal and along the transverse directions. 

The first measure is the number of atom layers that are involved in atomic mixing. If there are two layers of Si and two layers of Ge atoms that are involved in atomic intermixing, the number of atoms that are swapped per layer follows a pattern of 1|2|2|1. We label such interface structure by 4 ml, in short for four
mixing layers. A larger ml number corresponds to the larger degree of disorders in the cross-plane direction. The atomic number density of Si across the rough interface with different mixing layers can be found in Fig.~\ref{swap} in the Appendix.

The second measure is the size of the transverse supercell. For example, when we construct a 2$\times$2 transverse supercell with 4 ml structure, there are two out of four atoms for the Si atom layer closest to the interface and one out of four atoms for the Si atom layer secondly closest to the interface involved in atomic mixing. When we use a larger transverse supercell (3$\times$3 or 4$\times$4), we let the number of swapped atoms unchanged. The larger transverse supercell we use, the smaller degree of disorders along the transverse directions.
For a given ml number and a given supercell size, we generate 21 random configurations and compute the ensemble average of the transmission and reflection probability matrix elements.

\begin{figure}[t]
    \includegraphics[width=0.5\textwidth]{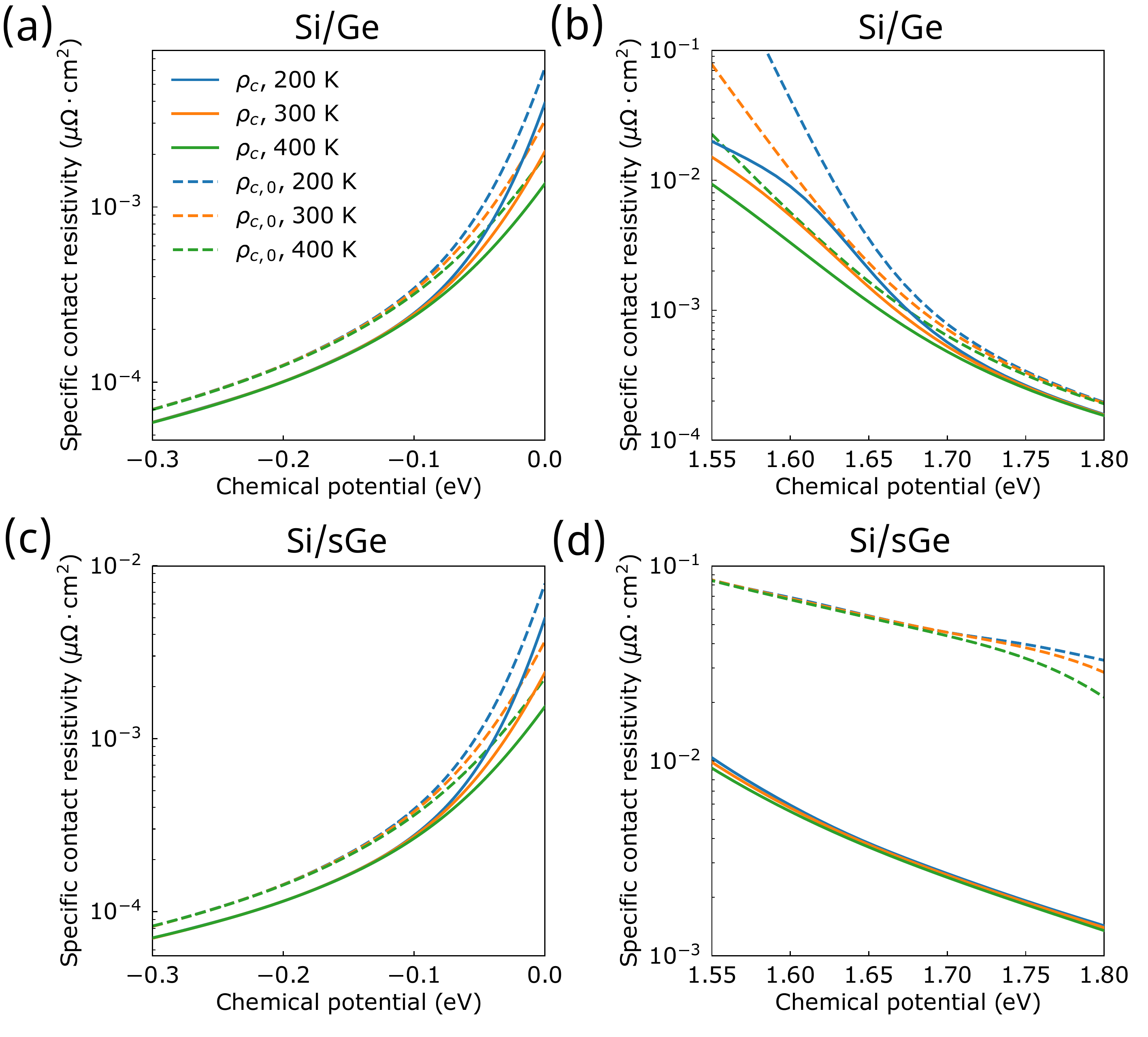}
    \caption{The specific contact resistance for (a) electrons and (b) holes for rough Si/Ge interfaces in solid lines and for the perfect Si/Ge interface in dashed lines at various temperatures. The specific contact resistance for (c) electrons and (d) holes for rough Si/sGe interfaces in solid lines and the perfect Si/sGe interface in dashed lines at various temperatures.}
    \label{resist}
\end{figure}

We found that the total transmissions for  electrons and holes are both enhanced for rough Si/Ge and Si/sGe interfaces compared with the corresponding perfect interfaces, as shown in Figs.~\ref{ns} (a)-(d). Particularly,
the total transmission function for a perfect Si/sGe interface with energy $E$ ranging from 1.5 to 1.9 eV is zero, whereas the transmission function for the corresponding rough interface is largely enhanced by nonspecular scattering processes. 
For the hole transmission, we notice that the nonspecular part is much smaller than the specular part and the enhancement in the total transmission is not significant, although the transmission for both electrons and holes is enhanced by the rough interface compared with the perfect interface. This is because holes are close to $\mathbf{k}=0$ and have long wavelengths. A more detailed discussion on the enhancement of hole transmission can be found in Appendix~\ref{holetrans}. In Fig.~\ref{sen1} and Fig.~\ref{sen2}, we have shown the dependence of transmission function on the degree of disorders along transverse and longitudinal directions. It turns out that the smaller transverse supercell dimensions and large mixing layer numbers are in favor of the nonspecular transmission. The smaller transverse supercell provides a large $\mathbf{G}_\parallel$, which allows the transition between valleys with large momentum mismatch. The larger degree of disorders along the perpendicular direction can lower the lateral symmetry to a greater extent and provides more channels that are previously forbidden by symmetry. Moreover, the effective thickness $\delta$ of the interface roughness along the perpendicular direction increases with increasing ml number. The interface roughness preferably couples with carriers with $|k_z| \sim \frac{2\pi}{\delta}$. The enhancement of transmission will be promoted if the corresponding valley satisfies $|k_z| \sim \frac{2\pi}{\delta}$.

In Fig.~\ref{ns} (e), we plot the mode-resolved specular transmission and reflection probabilities at $E$ = 1.77 eV as a function of their unfolded momentum $\mathbf{k}_{\mathrm{uc},\parallel}$. We find that the overlapped valleys for Si and sGe at the $\bar{\Gamma}$ point lead to small specular transmission probability. This is because the atomic mixing at the interface breaks the symmetry of Hamiltonian $\hat{H}$ and the above-mentioned symmetry-forbidden transmission at the $\bar{\Gamma}$ point is now allowed. In Fig.~\ref{ns} (f), we show the nonspecular transmission and reflection probabilities. The majority of nonspecular transmission processes are found to be starting from the $\bar{\Gamma}$ and $\bar{\mathrm{M}}$ points in Si to the $\bar{\mathrm{X}}$ and  $\bar{\mathrm{Y}}$ points in sGe. These processes correspond to the transition between the lowest conduction band of Si at the $\Delta$ point and the lowest conduction band of Ge at the $\mathrm{L}$ point in the three-dimensional Brillouin zone. 
Si's conduction band at the $\Delta$ point and Ge's conduction band at the $L$ point are both far from the $\Gamma$ point and the conduction electrons have small wavelengths. The characteristic length of disorders has to be small to contribute to the nonspecular interface scattering. Thus, smaller transverse supercell dimensions, \textit{i.e.}, atomic-scale disordered structures, are in favor of more nonspecular transmission channels.

We can define the specular and nonspecular reflection probabilities similarly to the transmission. By examining the specular and nonspecular reflection probabilities, we find that the newly emerged nonspecular reflection channels are accompanied by the removal of the specular transmission channels at the same $\mathbf{k}_{\mathrm{uc},\parallel}$. Although the increasing nonspecular reflection probability is detrimental for interfacial transport, there are overall more nonspecular transmission channels than the nonspecular reflection channels, thus, the total transmission is still enhanced.




With the knowledge of the transmission function, we proceed to compute the contact resistance.
The Landauer-B\"{u}ttiker formalism is used to compute the two-probe conductance,
\begin{equation}
     G_{12} = - \frac{2e^2}{h}\int dE\Theta(E)\frac{\partial f}{\partial E}
\end{equation}
where $h$ is the Plank constant, $f=\frac{1}{e^{(E-\mu)/k_BT}+1}$ is the Fermi-Dirac distribution function, and the factor 2 describes the spin degree of freedom. The subscripts 1 and 2 refer to the left and right sides. The four-probe conductance can be computed by~\cite{ferry1999transport,PhysRevB.23.6851},
\begin{equation}
     G_4 = \frac{1}{G^{-1}_{12}-\frac{1}{2}\left(G^{-1}_{11}+G^{-1}_{22}\right)} 
\end{equation}
where $G_{11}$ and $G_{22}$ are the two-probe conductance for bulk material $1$ and $2$, respectively. In practical calculations of the conductance for a bulk material, we let the two leads and devices all consist of the same materials. Then, the specific contact resistance is defined by,
\begin{equation}
\rho_c=\frac{A}{G_4}
\end{equation}
where $A$ is the cross-section area.


\begin{figure}
    \includegraphics[width=0.5\textwidth]{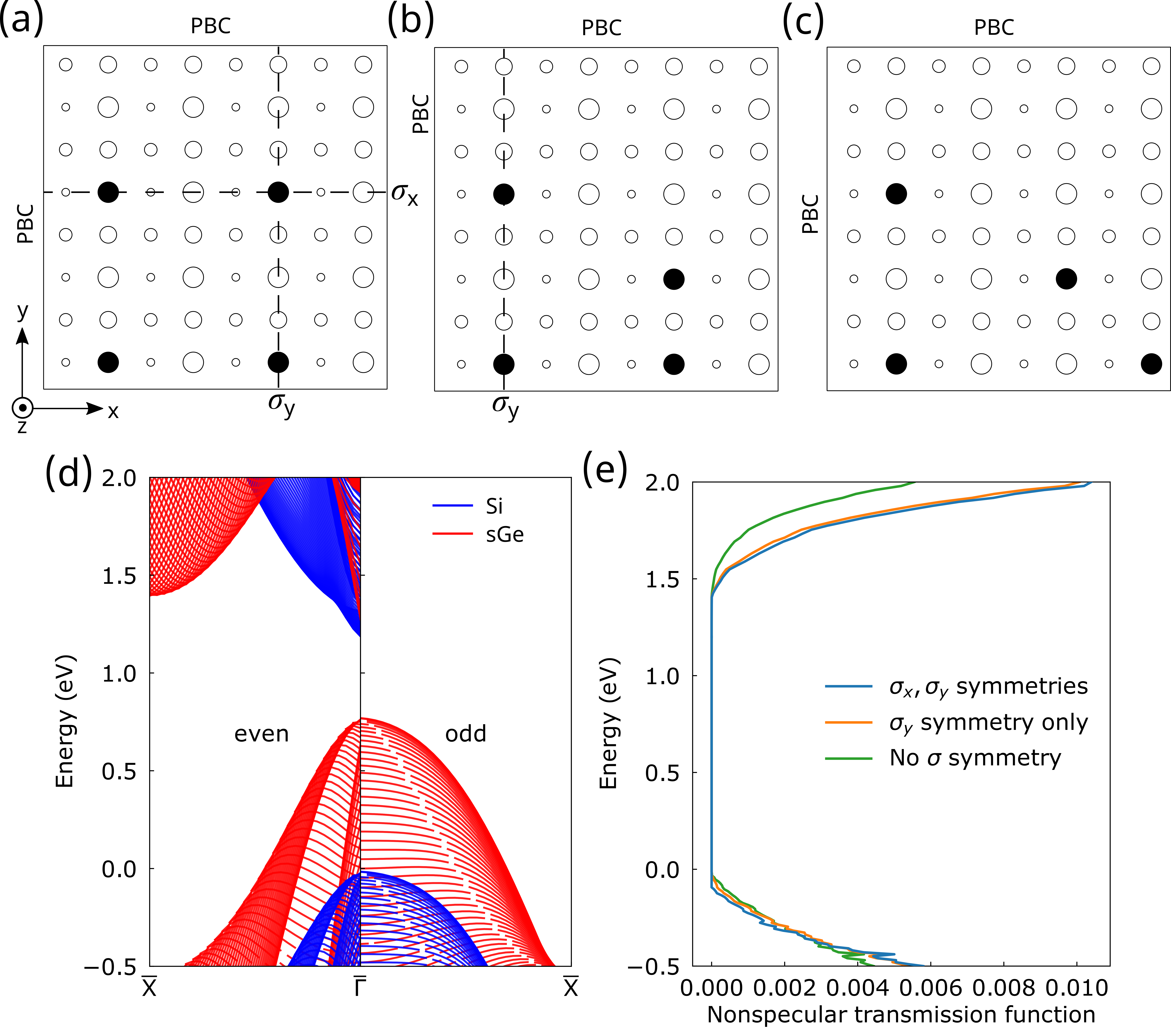}
    \caption{(a) The disordered structure with both $\sigma_x$ and $\sigma_y$ symmetry. (b) The disordered structure without $\sigma_x$ yet with $\sigma_y$ symmetry. (c) The disordered structure without $\sigma_x$ and $\sigma_y$ symmetry. $\sigma_x$ and $\sigma_y$ refer to the mirror symmetries along the x and y directions. In (a)-(c), the open circles are the Si atoms and the filled circles refer to the swapped Ge atoms. The larger circles represent the atoms closer to the interface. We only plot the Si side here, and on the Ge side the swapped Si atoms have the same transverse positions with the swapped Ge atoms on the Si side. (a)-(c) All correspond to $4\times4$, 2-ml structures. (d) The projected bulk band structures of Si and sGe along the $\bar{\mathrm{\Gamma}}\bar{\mathrm{X}}$ in the surface Brillouin zone sorted by their symmetries. (e) The nonspecular transmission function $\Theta_\mathrm{ns}(E)$ for the three disordered interface structures plotted in (a)-(c).}
    \label{sym}
\end{figure}


In Fig.~\ref{resist}, we observe over an order of magnitude reduction in the contact resistance for conduction electrons through the rough Si/sGe interface at various temperatures compared with the corresponding perfect interface. For the  rough Si/Ge interface, a much smaller reduction in electron contact resistance is found. This is because the perfect Si/Ge interface does not have the symmetry-forbidden transmission for low-energy conduction electrons as the perfect Si/sGe interface does. As a result, for Si/Ge interface, the lowered symmetry due to interface roughness does not benefit as much as the Si/sGe interface. For valence bands, the hole contact resistances for Si/Ge and Si/sGe interfaces are only slightly reduced by the interface disorders, as most holes from two sides have compatible momenta and symmetries. 

Last but not least, we want to examine how the symmetry of the disordered interface changes the nonspecular transmission.
The nonspecular transmission probability can be analyzed using perturbation theory~\cite{brataas1994semiclassical,PhysRevB.104.085310} and we argue that the nonspecular transmission probability is proportional to the scattering matrix element, $t_{ns,ji}\propto \bra{u_j}\Delta H\ket{u_i}$, where the perturbed potential is the difference between the potential energy for disordered interface and perfect interface, $\Delta H = H_\mathrm{rough}-H_\mathrm{perfect}$. For different disordered interface structures, the symmetry of $\Delta H$ can be different.

In Fig.~\ref{sym}, we show the nonspecular transmission for three representative disordered interface configurations. In Fig.~\ref{sym} (d), we have plotted the projected band structures of Si and sGe, sorted according to the symmetries of Bloch wave functions under mirror operation. The conduction band for Si and sGe are both even under the $\sigma_x$ operation, thus, it is preferred to have $\Delta H$ with even symmetry as well such that $t_{\mathrm{ns},ji}\propto \bra{u_j}\Delta H\ket{u_i}=\bra{\sigma_x u_j}\sigma_x\Delta H\ket{\sigma_xu_i} = \bra{u_j}\Delta H\ket{u_i}$ and $t_{\mathrm{ns},ji}$ is not forbidden by symmetry. As for the case with no mirror symmetries along $x$ or $y$ directions, the symmetry of the whole system is lowered and the symmetry analysis for $t_{\mathrm{ns},ji}$ does not work. Although there are some nonspecular transmission channels for the case with no mirror symmetry, the nonspecular transmission still favors the disordered structures with compatible symmetries with the initial and final states than those without.




In general, our symmetry analysis for wave functions and disordered structures can be applied to study other interfaces between semiconductors with mismatched band valleys. The extent of the contact resistance reduction depends on the specific materials on two sides of the interface and can only be known from Green's function calculations. However, the computational cost of Green's function calculation increases rapidly with number of atoms. On top of that, when the material is polar, the band edge profile near the interface (especially for complex oxide interfaces~\cite{PhysRevX.7.011023}) can vary significantly over a long distance, thus the Poisson equation has to be solved using a large supercell. Due to these challenges, we only study Si/Ge interfaces in this work.   


\section{Conclusion}\label{conclu}
We have studied the charge transport through a [001] Si/Ge interface. The transmission though a perfect interface must be specular. The electron transmission through the Si/Ge interface is very low due to momentum-mismatched band structures. The incompatible symmetries of the electron states at different pockets also forbid the transmission, leading to a high contact resistance. However, with atomic mixing at the interface, the symmetry is lowered and the previously forbidden transmission is allowed. In addition, the nonspecular transmission connecting electron pockets with different transverse momentum is enabled by those interface disorders. As a result, the specific contact resistance is reduced by over an order of magnitude.

\section{Acknowledgements}
We thank G.D. Mahan for helpful discussions.
This work is partially supported by the MRSEC Program of the National Science Foundation
under Award No. DMR-1419807.
G.C. gratefully acknowledges MIT support.


\appendix
\renewcommand{\thefigure}{A\arabic{figure}}
\setcounter{figure}{0}
\section{Mode-resolved Green's function formalism}\label{mgf1}

The mode-resolved Green's function formalism to compute the transmission and reflection probability matrix  is developed by Khomyakov \textit{et al.}. We present a brief introduction to the formalism as follows for completeness.
We first construct the Hamiltonian for the structure shown in Fig.~\ref{gf} (a). 
For a given $\mathbf{k}_{\mathrm{sc},\parallel}$, the Hamiltonian writes,
\begin{equation}
    H(\mathbf{k}_{\mathrm{sc},\parallel})=\begin{pmatrix}
        \ddots&&&&\\
       & H^L_{11} & H^L_{01}&&&\\
       & H^L_{10} &H^L_{00}&H_{LD}&&\\
       &&H_{DL}&H_D&H_{DR}&&\\
       &&&H_{RD}&H^{R}_{00}&H^{R}_{01}&\\
       &&&&H^R_{10}&H^R_{11}&\\
       &&&&&&\ddots
    \end{pmatrix}
\end{equation}
with the matrix blocks corresponding to different cells of the supercell along the interface normal as well as the interactions between neighboring cells.
In the semi-infinite lead region, we have $H^L_{nn}  = H^L_{00} $ and $H^R_{nn}  = H^R_{00} $, where $n$ denote the $n$th repeated supercell cell in the lead region as denoted in Fig. 1 (a). $H_{D}$ is the Hamiltonian corresponding to the device region. $H_{LD/DL}$ and $H_{RD/DR}$ describe interactions between the lead and the device region.

The Green's function matrix is defined by,
\begin{equation}
    \left[(E \pm i\eta) I-H(\mathbf{k}_{\mathrm{sc},\parallel})\right]G^{r/a}(E,\mathbf{k}_{\mathrm{sc},\parallel}) = I
\end{equation}
where $I$ is the identity matrix, $\eta$ is an infinitesimal positive real number, and the superscripts $r$ and $a$ denote retarded and advanced Green's function, depending on the sign in front of $i\eta$. The retarded Green's function at given energy $E$ in the block-matrix form is explicitly expressed by, 
\begin{equation}
    G^r(E,\mathbf{k}_{\mathrm{sc},\parallel})=\begin{pmatrix}
        \ddots&&&&&&\\
       & G^L_{11} & G^L_{01}&&&\udots\\
       & G^L_{10} &G^L_{00}&G_{LD}&G_{0,N+1}&\\
       &&G_{DL}&G_D&G_{DR}&&\\
       &&G_{N+1,0}&G_{RD}&G^{R}_{00}&G^{R}_{01}&\\
       &\udots&&&G^R_{10}&G^R_{11}&\\
       &&&&&&\ddots
    \end{pmatrix}
    \label{fullG}
\end{equation}

In particular, the matrix blocks $G^L_{00}$ and $G^R_{00}$ are useful physical quantities to compute the transmission matrix, called the surface Green's function for the left and right lead which satisfies,
\begin{equation}
g^{r}_{L/R}(E,\mathbf{k}_{\mathrm{sc},\parallel}) = G^{L/R}_{00} = [(E + i\eta)I - H^{L/R}_{00} - \Sigma^r_{L/R}]^{-1} 
\label{surfaceG}
\end{equation}
where the self-energies of the left and right leads are $\Sigma^r_{L} = H^L_{10} g^r_{L} H^L_{01} $ and $\Sigma^r_{R} = H^R_{01} g^r_{R} H^R_{10} $. In this work, the surface Green's functions are iteratively solved using the decimation 
technique~\cite{sancho1985highly}.

The matrix block $G_{N+1,0}$ describes the response at $N+1$ th cell ($0$ th cell in the right lead) cell given the perturbation at $0$ th cell in the left lead. We use Dyson's equation to compute the matrix $G_{N+1,0}$. To obtain $G_{N+1,0}$, we need to evaluate the matrix block $G_D$ for the device region, which contains a large amount of atoms for the case of rough interfaces. The efficient computation for device Green's function $G_D$ is thus challenging. To overcome this difficulty, the device Green's function is calculated using the recursive technique~\cite{Lewenkopf2013,Sols1989} and the detail for our implementation of the recursive Green's function can be found in our prior work~\cite{PhysRevB.104.085310}.
 
To compute the transmission mentioned in Eq.~\ref{trl}, we need to compute the eigenvector matrices and velocity matrices. We outline how to compute these matrices in the following.
For a given transverse momentum $\mathbf{k}_{\mathrm{sc},\parallel}$ and energy $E$, there are multiple subbands in the lead region with different perpendicular momenta $k_z$'s. What is more, the lead is semi-infinite, which supports both propagating (real $k_z$) and evanescent (imaginary $k_z$) states. We need to resolve the perpendicular momentum $k_z$ and its corresponding velocity $v_z$ to compute the ratio of scattered current to the incident current to obtain the transmission and reflection probability matrix.

We first introduce an auxiliary matrix for the right lead,
 \begin{equation}
    F^r_{R} = g^r_{R}H^R_{10}
    \label{Fmat}
 \end{equation}
and compute its eigenvalue $\Lambda^r_R$ and eigenvector $U^r_R$ via,
\begin{equation}
    F^r_{R}U^r_R=  \Lambda^r_R U^r_R 
    \label{eigen}
\end{equation}
It has been pointed out by Khomyakov \textit{et al.}~\cite{PhysRevB.72.035450} that the eigenvalue $\Lambda^r_{R,i}$ stores the phase information of the electron and the eigenvector matrix $U^r_{R,i}$ contains the Bloch wave functions for state $i$. If $|\Lambda^r_{R,i}| \neq 1$, it corresponds to an evanescent state. If $|\Lambda^r_{R,i}| = 1$, it corresponds to a propagating state. We can  extract the perpendicular momentum by $k_{R,i} = \frac{1}{a_R}\mathrm{log}\Lambda^r_{R,i}$. Similarly, for the left lead, we define the auxiliary matrix, and its eigenvalues and eigenvectors write,
 \begin{equation}
    F^a_{L} = g^a_{L}H^L_{01}
\end{equation}
\begin{equation}
    F^a_{L}U^a_L = \Lambda^a_L U^a_L
\label{Fmat1}
\end{equation}
where $g^a_{L}=(g^r_{L})^\dagger$ is the advanced surface Green's function for the left lead.

The velocity along the transport direction (perpendicular to interface) $v_z$ can  be described by the velocity matrix,
\begin{equation}
    V^a_{L} = -U^{a\dagger}_{L}\Gamma^a_LU^{a}_{L}
\end{equation}
\begin{equation}
    V^r_{R} = U^{r\dagger}_{R}\Gamma^r_RU^{r}_{R}.
    \label{velmat}
\end{equation}
where $\Gamma = i(\Sigma- \Sigma^{\dagger})$.
The diagonal elements of these matrices correspond to the group velocities along z direction of different states. 

The reflection probability matrix from $\alpha$ side $R_{\alpha\alpha,ji}(E,\mathbf{k}_{\mathrm{sc},\parallel})$
is similarly defined by,
\begin{equation}
     R_{\alpha\alpha,ji}(E,\mathbf{k}_{\mathrm{sc},\parallel}) = |r_{\alpha\alpha,ji}(E,\mathbf{k}_{\mathrm{sc},\parallel})|^2
\end{equation}
Specifically, the reflection matrices from the left and right sides are,
\begin{equation}
     r_{LL}(E,\mathbf{k}_{\mathrm{sc},\parallel})  = i\sqrt{V^r_L}[U^r_L]^{-1}\left(G_{0,0}-Q^{-1}_L\right)[U^{a\dagger}_L]^{-1}\sqrt{V^a_{L}}
\label{refl0}
\end{equation}
\begin{equation}
     r_{RR}(E,\mathbf{k}_{\mathrm{sc},\parallel})  = i\sqrt{V^r_R}[U^r_R]^{-1}\left(G_{N+1,N+1}-Q^{-1}_R\right)[U^{a\dagger}_R]^{-1}\sqrt{V^a_{R}}
\label{refl}
\end{equation}
where $Q^{-1}_L = (E+i\eta)I - H^L_{00}-H^{L}_{10}g^r_LH^{L}_{01}-H^{L}_{01}g^r_{L'}H^{L}_{10}$ and $Q^{-1}_R = (E+i\eta)I - H^R_{00}-H^{R}_{01}g^r_RH^{L}_{10}-H^{R}_{10}g^r_{R'}H^{R}_{01}$ are the retarded Green's functions for bulk materials. $g^r_{\alpha'},\;\alpha=L,R$, are the retarded surface Green's function similar to Eq.~\ref{surfaceG}, except that they describe the semi-infinite lead of the same material extending to infinity in the opposite direction given by,
\begin{equation}
g^r_{L^\prime}(E,\mathbf{k}_{\mathrm{sc},\parallel})=\left[(E+i\eta)I-H^L_{00}-\Sigma^r_{L^\prime}\right]^{-1}
\end{equation}
\begin{equation}
g^r_{R^\prime}(E,\mathbf{k}_{\mathrm{sc},\parallel})=\left[(E+i\eta)I-H^R_{00}-\Sigma^r_{R^\prime}\right]^{-1}\\
\end{equation}
where the self-energies write,
\begin{equation}
\Sigma^r_{L^\prime}=H^L_{01}g^r_{L^\prime}H^L_{10}
\end{equation}
\begin{equation}
\Sigma^r_{R^\prime}=H^R_{10}g^r_{R^\prime}H^R_{01}
\end{equation}

The reflection matrix also depends on another two surface Green's function $g^r_L$ and $g^a_R$, as defined by Eq.~\ref{surfaceG}.
The auxiliary matrices, eigenvalue matrices, and eigenvector matrices for these two surface Green's functions are,
\begin{equation}
F_L^r=g^r_L H^L_{01}
\end{equation}
\begin{equation}
F_L^rU_L^r=\Lambda_L^rU_L^r
\end{equation}
\begin{equation}
F_R^a=g^a_R H^R_{10}
\end{equation}
\begin{equation}
F_R^aU_R^a=\Lambda_R^aU_R^a
\end{equation}
The self-energies for these two surface Green's functions are,
\begin{equation}
\Sigma_R^a = H^R_{01}g^r_RH^R_{10}
\end{equation}
\begin{equation}
\Sigma_L^r = H^L_{10}g^a_LH^L_{01}
\end{equation}
The corresponding broadening matrices are computed by $\Gamma = i(\Sigma-\Sigma^\dagger)$.
The velocity matrices $V^a_R$ and $V^r_L$ introduced in Eq.~\ref{refl0} and Eq.~\ref{refl} are expressed by,
\begin{equation}
        V^a_{R} =- U^{a\dagger}_{R}\Gamma^a_{R}U^{a}_{R}
\end{equation}
\begin{equation}
        V^r_{L} = U^{r\dagger}_{L}\Gamma^r_{L}U^{r}_{L}
\end{equation}

\section{Interface atomic mixing}
\begin{figure}
    \includegraphics[width=0.28\textwidth]{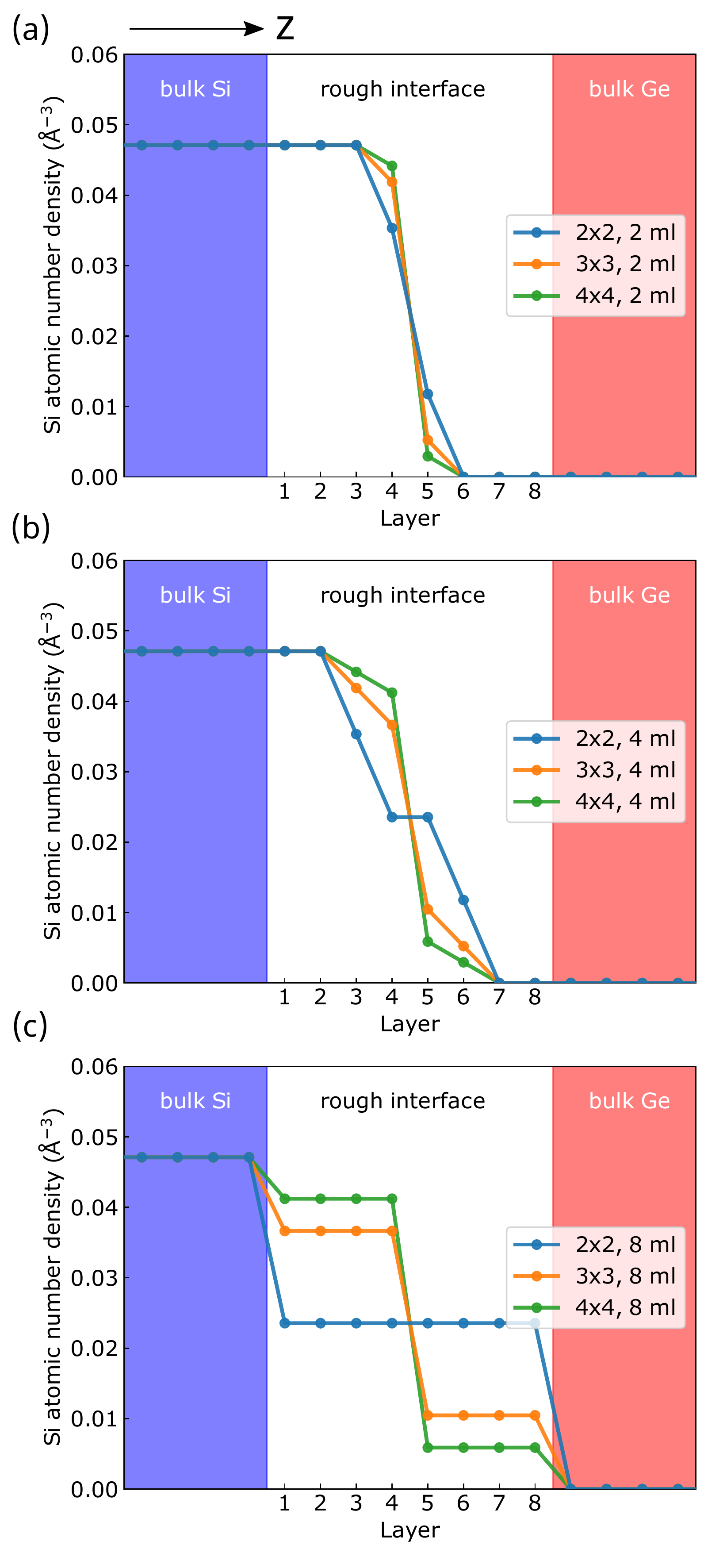}
    \caption{The atomic number density (the number of atoms of a given type per unit volume) of Si at different atom layers of the interface.}
    \label{swap}
\end{figure}

\begin{figure}
    \includegraphics[width=0.5\textwidth]{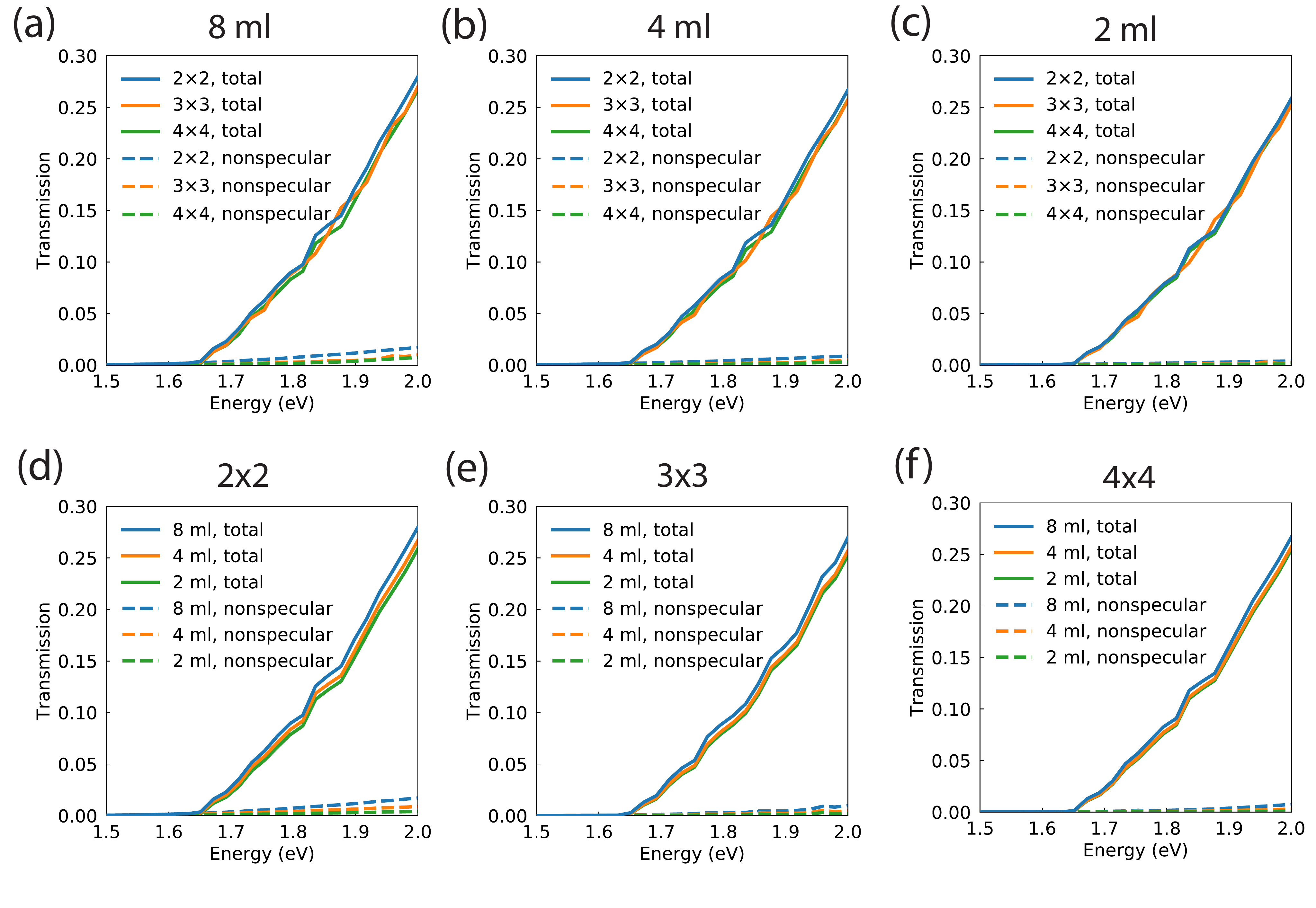}
    \caption{The sensitivity of electron transmission through Si/Ge interfaces on the degree of transverse and longitudinal disorders. In (a)-(c), we use a 2$\times$2 transverse supercell and change the ml number. In (d)-(f), we use 8 ml structures and vary the in-plane supercell size. We use 20$\times$20, 15$\times$15, 10$\times$10 $\mathbf{k}_{\mathrm{sc},\parallel}$-point mesh for 2$\times$2, 3$\times$3, 4$\times$4 transverse supercells, respectively.}
    \label{sen1}
\end{figure}

\begin{figure}
    \includegraphics[width=0.5\textwidth]{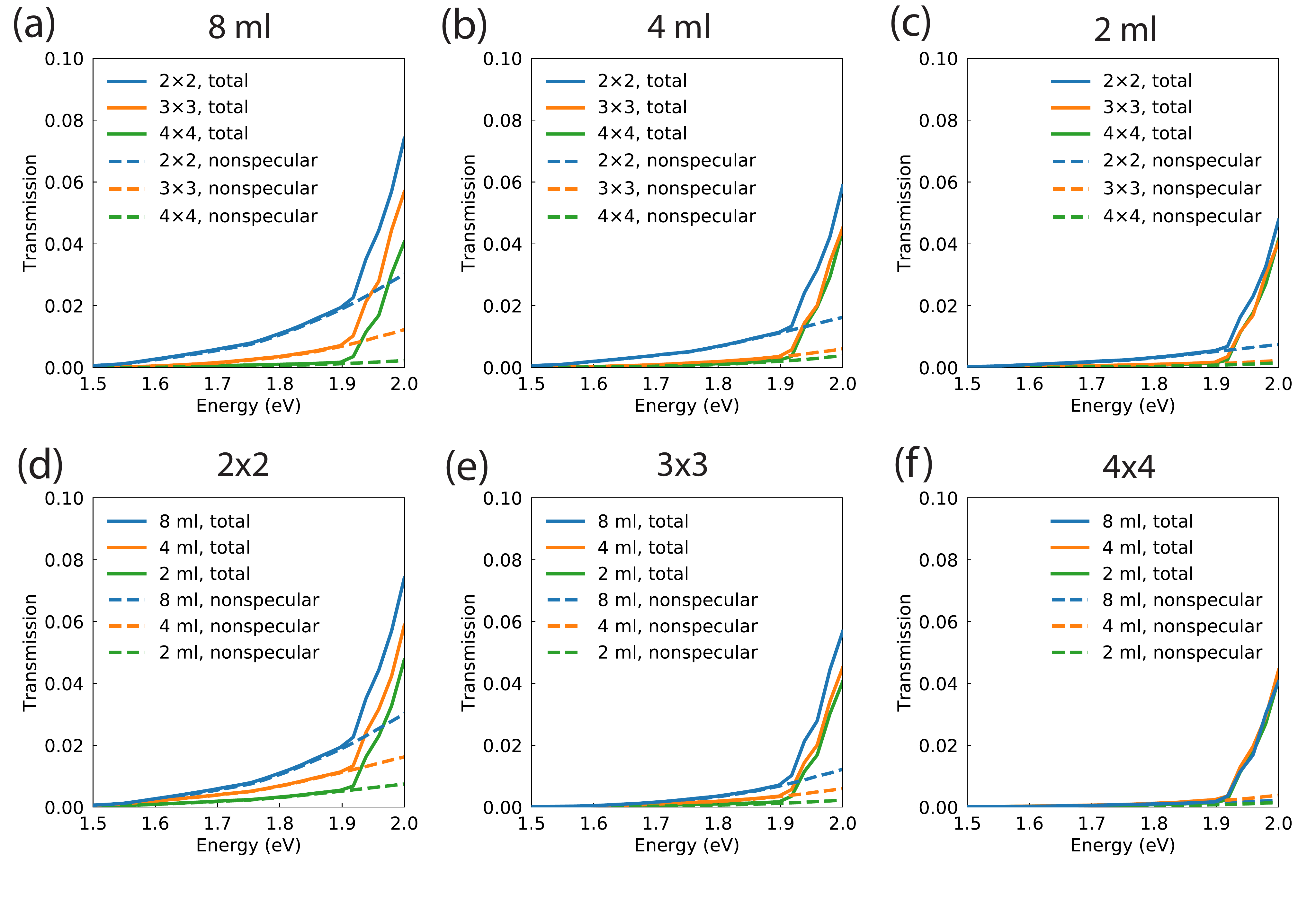}
    \caption{The sensitivity of electron transmission through Si/sGe interfaces on the degree of transverse and longitudinal disorders. Other computational details are the same with those for Fig.~\ref{sen1}.}
    \label{sen2}
\end{figure}

In Fig.~\ref{swap}, we present the ensemble-averaged  atomic number density along \textit{z} direction for different ml and transverse supercell sizes. In Fig.~\ref{sen1} and Fig.~\ref{sen2}, we demonstrate the corresponding electron transmission for different interface configurations.  At fixed ml number, smaller transverse supercell sizes give rise to higher nonspecular transmission function. At fixed transverse supercell size, the higher ml numbers give rise to higher nonspecular transmission function.




\section{The enhancement of hole transmission}\label{holetrans}
\begin{figure}
    \includegraphics[width=0.5\textwidth]{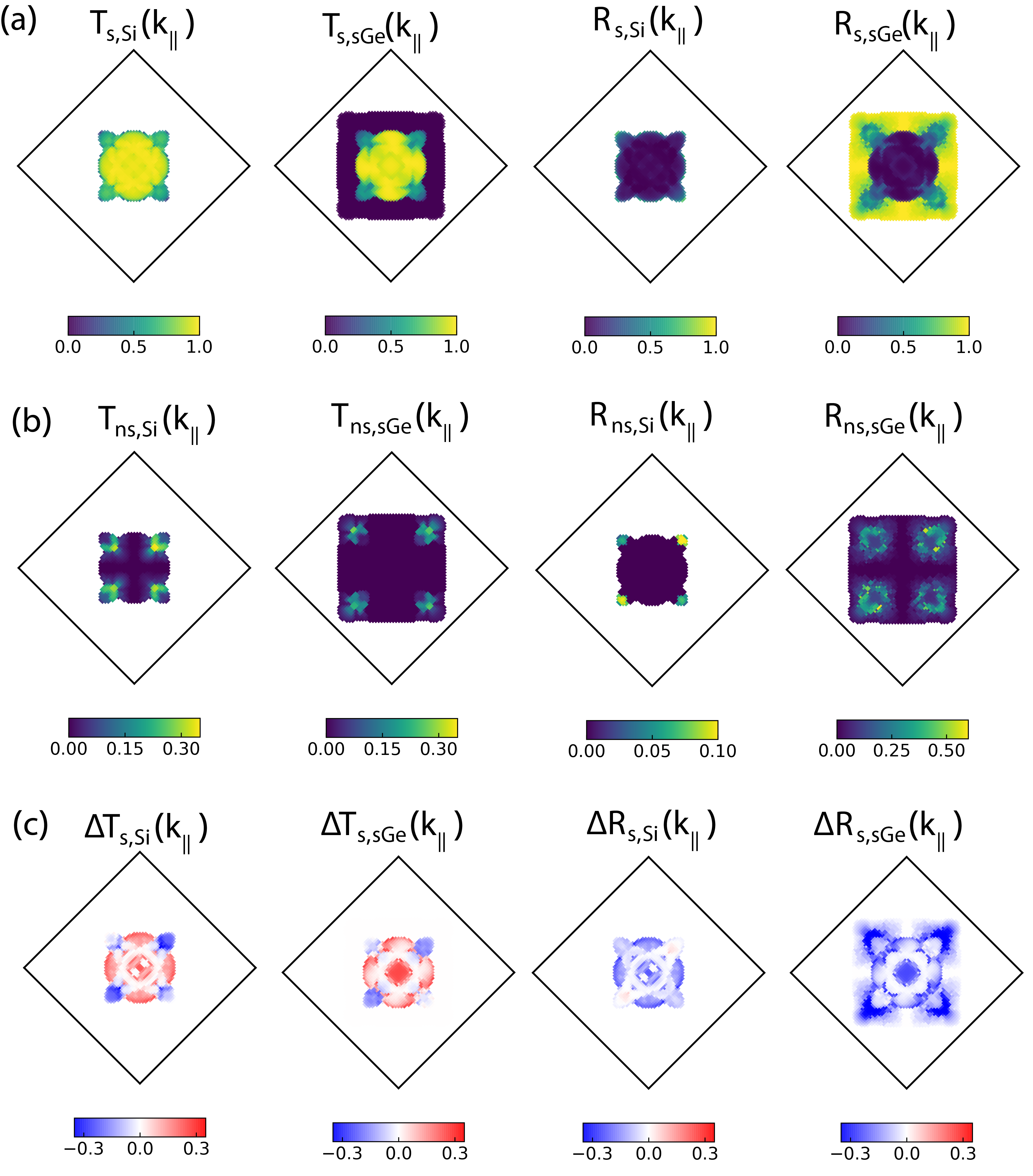}
    \caption{The ensemble-averaged mode-resolved (a) specular and (b) nonspecular scattering probabilities as a function of in-plane momentum $\mathbf{k}_{\mathrm{uc},\parallel}$ at $E = -0.41$ eV. (c) The difference between the ensemble-averaged specular scattering probability for disordered interfaces and the specular scattering probability for the perfect interface, $\Delta P_{\mathrm{s},\alpha}(\mathbf{k}_{\mathrm{uc},\parallel}) = P_{\mathrm{s},\alpha}(\mathbf{k}_{\mathrm{uc},\parallel})  - P_{\mathrm{perfect},\alpha}(\mathbf{k}_{\mathrm{uc},\parallel})$, where $P=T,R$ and $\alpha=\mathrm{Si,sGe}$, as a function of in-plane momentum $\mathbf{k}_{\mathrm{uc},\parallel}$ at $E = -0.41$ eV.}
    \label{enh_hole}
\end{figure}

From Fig.~\ref{ns}, we observe that the hole transmission functions for disordered Si/Ge and Si/sGe interfaces are both slightly enhanced compared with corresponding perfect interfaces. To investigate the origin of such enhancement, we compare the mode-resolved hole scattering probabilities for rough Si/sGe interfaces and the perfect Si/sGe interface in Fig.~\ref{enh_hole}.

First, we notice that the specular part of the scattering probability is generally much larger than the nonspecular part except for the reflectance from the Si side. This is because the valence bands of Si and sGe are both at the $\bar{\Gamma}$ point, which means that the momentum conservation is easier to satisfy. Also, the symmetries for hole wave functions from two sides are compatible, which puts no symmetry restriction on hole transmission and reflection. Second, the nonspecular scattering probabilities (both transmittance and reflectance) are promoted at the points with relatively large $|\mathbf{k}_\parallel|$ on the Si and sGe side, such that those points can differ by integer number of transverse reciprocal vectors $\mathbf{G}_{\mathrm{sc},\parallel}$. Furthermore, compared with the perfect interface, the specular transmittance for rough interfaces can be either reduced or enhanced, depending on the transverse momentum. In contrast, the specular reflectance for most holes from Si and sGe are reduced. Consequently, the total hole transmission is slightly enhanced. However, such enhancement might not be universal. The reasons for this are as follows.

For the perfect interface, there are already a considerable amount of scattering channels (all of them are specular). When interfacial disorders are introduced, the change in specular scattering probability $\Delta P_{\mathrm{s},\alpha}(\mathbf{k}_{\mathrm{uc},\parallel})$ and the nonspecular scattering probability $P_{\mathrm{ns},\alpha}(\mathbf{k}_{\mathrm{uc},\parallel})$ are both small perturbations compared in the specular scattering probability for the perfect interface $P_{\mathrm{perfect},\alpha}(\mathbf{k}_{\mathrm{uc},\parallel})$. The signs of these perturbations depend on both the wave functions of the initial and final states. Specifically, it is difficult to predict how the specular part of the scattering probability changes using symmetry analysis. Eventually, the scattering probability for rough interfaces $P_{\alpha}(\mathbf{k}_{\mathrm{uc},\parallel})=P_{\mathrm{s},\alpha}(\mathbf{k}_{\mathrm{uc},\parallel})+P_{\mathrm{ns},\alpha}(\mathbf{k}_{\mathrm{uc},\parallel})$ can be either higher or lower than the scattering probability for the perfect interface $P_{\mathrm{perfect},\alpha}(\mathbf{k}_{\mathrm{uc},\parallel})$.

\clearpage
\bibliographystyle{apsrev4-2}
\bibliography{main_v4} 

\begin{thebibliography}{64}
\bibitem [{bon()}]{bond}%
  \BibitemOpen
  \href@noop {} {\bibinfo  {journal} {We are interested in the interface
  scattering by the atomically rough interfaces. Hence, we focus on the domain
  within the range of $\sim$nm to the interface, where the lattice constant
  does not vary drastically. Thus, we assume the Si-Si, Ge-Ge bond lengths are
  the same with Si-Ge bond length of 2.3979 \AA$\,$ in the domain of our
  simulation. That is to say, the Si/Ge interface structure is the same with
  bulk Si's structure except some atoms are substituted by Ge atoms}\
  }\BibitemShut {NoStop}%
\bibitem [{off()}]{offset}%
  \BibitemOpen
\bibfield  {journal} {  }\href@noop {} {\bibinfo  {journal} {The band structure
  transits from Si to Ge with a discontinuity of valence band edge known as the
  valence band offset. In addition, depending on the carrier concentrations of
  Si and Ge side, the band structures of Si and Ge vary with distance to the
  interface known as band bending. The characteristic length of band bending is
  the Debye length, which is typically around $\sim$10 to $\sim$100 nm. Since
  we focus on the interface scattering in the proximity of rough interfaces
  (below 10 nm), we neglect the band bending and assume a constant band edge
  profile in Si side and Ge side. As for the band offset across the Si/Ge
  interface, we use the value of 0.68 eV for bond length $L = L_0 = 2.398$
  \AA}\ }\BibitemShut {NoStop}%
\end{thebibliography}
\end{document}